\documentclass[fleqn,twoside]{article}
\usepackage{espcrc2}
\usepackage{graphicx}
\usepackage{bm}

\newcommand{\AmS}{{\protect\the\textfont2
  A\kern-.1667em\lower.5ex\hbox{M}\kern-.125emS}}
\newcommand{\pvec}{\bm{p}}
\newcommand{\xvec}{\bm{x}}
\title{Baryonic Operators for Lattice Simulations}

\author{LHP Collaboration: R.~Edwards\address[JLAB]{%
                        Thomas Jefferson National Accelerator Facility,
                        Newport News, VA 23606, USA},
        R.~Fiebig\address[FIU]{%
                        Physics Department, 
                        Florida International University, 
                        Miami, FL 33199, USA},
        G.~Fleming\addressmark[JLAB],
        U.M.~Heller\address{%
                        American Physical Society,
                        One Research Road,
                        Ridge, NY 11961-9000, USA},
        C.~Morningstar\address[CMU]{%
                        Department of Physics, 
                        Carnegie Mellon University, 
                        Pittsburgh, PA 15213, USA}
                 \thanks{Poster presented by C.~Morningstar.},
        D.~Richards\addressmark[JLAB],
        I.~Sato\address[UMD]{%
                        Department of Physics, 
                        University of Maryland,
                        College Park, MD 20742, USA},
        S.~Wallace\addressmark[UMD].
        }
       
\begin{document}

\begin{abstract}
The construction of baryonic operators for determining the $N^\ast$
excitation spectrum is discussed.  The operators are designed with one
eye towards maximizing overlaps with the low-lying states of interest,
and the other eye towards minimizing the number of sources needed in
computing the required quark propagators.  Issues related to spin
identification are outlined.  Although we focus on
tri-quark baryon operators, the construction method is applicable
to both mesons and penta-quark operators.
\end{abstract}
\maketitle

\section{OVERVIEW}

One major goal of the Lattice Hadron Physics Collaboration (LHPC)
is to calculate a significant portion of the low-lying spectrum of
baryon resonances from Monte Carlo estimates of Euclidean-space path
integrals in lattice QCD.  To accomplish this goal, large sets of operators
which can create the baryon states of interest at a given time $t$ must be
designed.  Our approach to constructing such operators is outlined
in this poster.

Baryon states are identified by their momentum $\pvec$,
intrinsic (half-integral) spin $J$,
projection $\lambda$ of this spin onto some axis,
parity $P=\pm 1$, and
quark flavor content (isospin, strangeness, {\it etc.}).
Since we are interested first in the masses of these states, we
restrict our attention to the $\pvec=\bm{0}$ sector, so our
operators must be invariant under all spatial translations
allowed on a cubic lattice.  The little group of all symmetry
transformations which leave $\pvec=\bm{0}$ invariant is the
octahedral point group $O_h$, so our operators may be classified
using the spinorial irreducible representations (irreps) of $O_h$.
There are four two-dimensional irreps $G_{1g}, G_{1u}, G_{2g}$, $G_{2u}$
and two four-dimensional representations $H_g$ and 
$H_u$\cite{johnson,mandula2}. 
The continuum-limit spins $J$ of our states must be deduced by examining
degeneracy patterns across the different $O_h$ irreps.

We use a two-step approach to construct the operators.
First, we design {\em elemental} operators $B^{F}_i(t,\bm{x})$
having the appropriate flavor structure characterized by isospin, strangeness,
{\it etc.}, and color structure constrained by gauge invariance.
In the second step, we apply group-theoretical projections to
obtain operators which transform irreducibly under 
all lattice rotation and reflection symmetries:
\begin{equation}
  B_i^{\Lambda\lambda F}\!(t,\bm{x})\!
 = \!\frac{d_\Lambda}{g_{O_h}}\!\!\!\sum_{R\in O_h}\!\!
  \!\!D^{(\Lambda)\ast}_{\lambda\lambda}(R)
   U_R B^F_i\!(t,\bm{x}) U_R^\dagger,
\label{eq:project}\end{equation}
where $\Lambda$ refers to an $O_h$ irrep, $\lambda$ is the irrep row,
$g_{O_h}$ is the number of elements in $O_h$,
$d_\Lambda$ is the dimension of the $\Lambda$ irrep,
$D^{(\Lambda)}_{mn}(R)$ is a $\Lambda$
representation matrix corresponding to group element $R$,
and $U_R$ is the quantum operator which implements the symmetry
operations. We also impose translation invariance
for zero momentum states:
\begin{equation}
  B^{\Lambda\lambda F}_i(t)=\sum_{{\bm{x}}} 
   B^{\Lambda\lambda F}_i(t,\xvec).
\end{equation}

Note that in order to capture the {\em spectrum},  we plan to
calculate correlation matrices
\begin{equation}
 C_{ij}^{\Lambda\lambda F}(t)=\langle 0\vert\ T B^{\Lambda\lambda F}_i(t)
 \overline{B}^{\Lambda\lambda F}_j(0)\ \vert 0\rangle.
\end{equation}
Matrices of correlation functions facilitate the reliable extraction
of the masses of excited states since variational techniques, possibly
combined with Bayesian methods, can be exploited.
The operators are designed with one
eye towards maximizing overlaps with the low-lying states of interest,
and the other eye towards minimizing the number of sources needed to
compute the required quark propagators.  Throughout this work,
we work in the approximation $m_u=m_d$, where $m_u$ and $m_d$ are
the masses of the $u$ and $d$ quarks, respectively, and we ignore
electromagnetic effects.  Hence, isospin is a good quantum number.

\section{DESIGN DETAILS}

Gauge invariance and isospin dictate the color and flavor
structure of the elemental operators in the first stage of
the operator construction.  Since we intend to include all
allowed operators within a given isospin channel, we do not need
to consider explicit $SU(3)$-flavor combinations.  As long as we
are careful to respect isospin, the variational method takes
care of finding the preferred flavor combinations for us.

Both link variable and quark field smearings are important for
reducing the couplings of our operators to higher-lying states.
Familiar APE smearing\cite{APEsmear} of the spatial links
can be employed, and the three-dimensional
covariant spatial Laplacian is used to smear the quark fields:
\begin{equation}
  \tilde{\psi}(x) = (1+\varrho \tilde{\Delta})^{n_\varrho}\ \psi(x),
\end{equation}
where $\varrho$ and integer $n_\varrho$ are tunable parameters
and the three-dimensional gauge-covariant Laplacian is defined by
\begin{equation}
   \tilde{\Delta} \psi(x) = \sum_{k=\pm 1, \pm 2,\pm 3} \Bigl(
   \tilde{U}_k(x)\psi(x\!+\!\hat{k})-\psi(x) \Bigr),
\end{equation}
with $\tilde{U}_\mu(x)$ denoting a smeared link variable.
Note that the square of the smeared field is equal to zero,
similar to the simple Grassmann field.

Our three-quark elemental operators are constructed using the
following building blocks:
\begin{eqnarray}
\chi^n_{Aa\alpha}(x)&\equiv& 
  \bigl( \tilde{\Delta}^{n} \tilde{\psi}(x)\bigr)_{Aa\alpha},\\
\xi^{npj}_{Aa\alpha}(x)&\equiv& \bigl(\tilde{D}^{(p)}_j\!\tilde{\Delta}^{n} 
           \tilde{\psi}(x)\bigr)_{Aa\alpha},
\end{eqnarray}
where $A$ is a flavor index, $a$ is a color index, $\alpha$ is a
Dirac spin index, and the $p$-link gauge-covariant displacement
operator is defined by
\begin{equation}
 \tilde{D}_j^{(p)}O(x) = 
  \tilde{U}_j(x)\dots 
   \tilde{U}_j(x\!+\!(p\!-\!1)\hat{j}) O(x\!+\!p\hat{j}).
\end{equation}
Our three-quark elemental operators have the following forms:
\begin{eqnarray}
 & & \phi^F_{ABC}\ \varepsilon_{abc} 
   \ \chi^{n_1}_{Aa\alpha} 
   \ \chi^{n_2}_{Bb\beta}
   \ \chi^{n_3}_{Cc\gamma}, \\
  & & \phi^F_{ABC}\ \varepsilon_{abc}
 \ \chi^{n_1}_{Aa\alpha}
 \ \chi^{n_2}_{Bb\beta}
 \ \xi^{n_3 pj}_{Cc\gamma}, \\
  & & \phi^F_{ABC}\ \varepsilon_{abc}
 \ \chi^{n_1}_{Aa\alpha} 
 \ \xi^{n_2 p_1 j}_{Bb\beta}
 \ \xi^{n_3 p_2 k}_{Cc\gamma},
\end{eqnarray}
where $n_1, n_2, n_3, p, p_1, p_2$ are positive integers,
$j,k=\pm 1,\pm 2, \pm 3$ are spatial directions,
$\alpha, \beta, \gamma$ are Dirac spin indices,
$A, B, C$ are flavors, and $a,b,c$ are colors.
The Levi-Civita symbol $\varepsilon_{abc}$ specifies the
color structure and the $\phi^F_{ABC}$ specify the flavor
combinations.
Since baryon resonances are expected to be large objects,
extended operators are crucial.  
Note that the different powers of the spatial Laplacian $\tilde{\Delta}$
are used to build up radial structure, while the
displacement operator $\tilde{D}_j^{(p)}$ is used to incorporate orbital
structure.  A large number of baryons can be studied using a 
somewhat small number of quark propagator sources:
$ n_\kappa ( n_{a}+3n_p n_{b})$, where
$n_p$ is the number of displacement lengths $p$,
$n_a$ is the number of powers of $\tilde{\Delta}$,
$n_b$ is the number of powers of $\tilde{\Delta}$
combined with a displacement, and
$n_\kappa$ is number of quark masses used.

In the second stage of the operator construction, the
group-theoretical projections in Eq.~(\ref{eq:project})
are applied.  Eq.~(\ref{eq:project}) requires explicit
representation matrices for every group element.
Representation matrices for all allowed proper rotations can be
generated from the matrices for $C_{4y}$ and $C_{4z}$, the
rotations by $\pi/2$ about the $y$- and $z$-axes, respectively.
The explicit matrices we use for these generators are given by
\[
 D^{(G_1)}(C_{4y})=\frac{1}{\sqrt{2}}\!\left[\begin{array}{rr}
 \!\!1 & \!\!\!\!-1\!\! \\ \!\!1 & \!\!\!\! 1\!\!
\end{array}\right]=-D^{(G_2)}(C_{4y}),
\]
\[
 D^{(G_1)}(C_{4z})=\frac{1}{\sqrt{2}}\!\left[\begin{array}{cc}
  1\!-\!i & \!\!\!\!0 \\ 0 &\!\!\!\! 1\!+\!i 
\end{array}\right]\!\! = -D^{(G_2)}(C_{4z}),
\]
\[
 D^{(H)}(C_{4y})=\frac{1}{2\sqrt{2}}\left[\begin{array}{rrrr}
  \!1 \!&\! -\sqrt{3} \!&\! \sqrt{3} & -1 \!\\
  \!\sqrt{3} \!&\! -1 \!&\! -1 \!&\!  \sqrt{3} \!\\
  \!\sqrt{3} \!&\!  1 \!&\! -1 \!&\! -\sqrt{3}\! \\
  \!1 \!&\!  \sqrt{3} \!&\! \sqrt{3} &  1\! \end{array}\right],
\]
\[
 D^{(H)}(C_{4z})=\frac{1}{\sqrt{2}}\left[\begin{array}{cccc}
   \!\!-1\!-\!i \!&\! 0 \!&\! 0 & 0 \!\!\\
   \!\!0 \!&\! 1\!-\!i \!&\! 0 & 0 \!\!\\
   \!\!0 \!&\! 0 \!&\! 1\!+\!i & 0 \!\!\\
   \!\!0 \!&\! 0 \!&\! 0 \!&\! -1\!+\!i \!\!\end{array}\right].
\]
Inclusion of spatial inversion is straightforward.

The projections in Eq.~(\ref{eq:project}) are carried out
using Maple.  The resulting operators transform irreducibly
under $O_h$.  In the continuum limit, the spin $J$ is identified
using Table~\ref{tab:spin}.  For example, to identify a
spin-$\frac{1}{2}$ baryon, it must occur in the $G_1$ irrep
and accompanying degenerate levels in the $G_2$ and $H$ irreps
must {\em not} occur.  A level occurring at the same
$a\rightarrow 0$ mass in all three $G_1$, $G_2$, and $H$ irreps
can be identified as a spin-$\frac{7}{2}$ baryon.

\begin{table}[t]
\caption{Continuum limit spin identification:
   occurrences $n_\Gamma^J$ of $O$ irrep $\Gamma$
  in subduction of $SU(2)$ irrep $J$.
\label{tab:spin}}
\begin{center}
\renewcommand{\tabcolsep}{1.5pc} 
\renewcommand{\arraystretch}{1.2} 
\begin{tabular}{lrrr}\hline
  $J$  & $n^J_{G_1}$ & $n^J_{G_2}$ & $n^J_{H}$ \\ \hline
  $\frac{1}{2}$  &  $1$ & $0$ & $0$ \\
  $\frac{3}{2}$  &  $0$ & $0$ & $1$ \\
  $\frac{5}{2} $ &  $0$ & $1$ & $1$ \\
  $\frac{7}{2} $ &  $1$ & $1$ & $1$ \\
  $\frac{9}{2} $ &  $1$ & $0$ & $2$ \\
  $\frac{11}{2}$ &  $1$ & $1$ & $2$ \\
  \hline
\end{tabular}
\end{center}
\end{table}

To illustrate our operator construction, consider an
explicit example.  Let
\begin{equation}
  \Phi_{\alpha\beta\gamma}=
  \varepsilon^{abc}\left( u^a_\alpha\ d^b_\gamma\ u^c_\beta 
  - d^a_\alpha\ u^b_\gamma\ u^c_\beta\right),
\label{eq:example}\end{equation}
ignoring smearing and powers of the Laplacian to simplify
matters and using $u,d$ to indicate flavor. 
The above operators have isospin $I=1/2$, 
isospin projection $I_3=1/2$, and strangeness $S=0$.
There are twenty independent operators due to the constraints
$\Phi_{\alpha\beta\gamma}+\Phi_{\gamma\beta\alpha}=0$ and
$\Phi_{\alpha\beta\gamma}+\Phi_{\beta\gamma\alpha}
 +\Phi_{\gamma\alpha\beta}=0$.
Linear combinations which transform irreducibly under $O_h$
are listed in Table~\ref{tab:results}.

\begin{table}[t]
\caption{Combinations of the operators $\Phi_{\alpha\beta\gamma}$
 in Eq.~(\protect\ref{eq:example}) 
 which transform irreducibly under $O_h$ for the Dirac-Pauli
 representation of the $\gamma$-matrices.
\label{tab:results}}
\begin{center}
\renewcommand{\tabcolsep}{4mm} 
\begin{tabular}{ccc}\hline
 Irrep & Row &  Operators
    \\ \hline 
  $G_{1g}$  & 1 &  $ \Phi_{221}                         $  \\
  $G_{1g}$  & 2 &  $ \Phi_{112}                         $  \\ \hline
  $G_{1g}$  & 1 &  $ \Phi_{324}                         $  \\
  $G_{1g}$  & 2 &  $ -\Phi_{314}                        $  \\ \hline
  $G_{1g}$  & 1 &  $ 2\Phi_{441}+2\Phi_{234}-\Phi_{324} $  \\ 
  $G_{1g}$  & 2 &  $ 2\Phi_{332}+2\Phi_{134}-\Phi_{314} $  \\ \hline
  $G_{1u}$  & 1 &  $ \Phi_{443}                         $  \\ 
  $G_{1u}$  & 2 &  $ \Phi_{334}                         $  \\ \hline
  $G_{1u}$  & 1 &  $ \Phi_{124}-\Phi_{214}              $  \\
  $G_{1u}$  & 2 &  $ -\Phi_{123}+\Phi_{213}             $  \\ \hline
  $G_{1u}$  & 1 &  $ 2\Phi_{223}-\Phi_{124}-\Phi_{214}  $  \\ 
  $G_{1u}$  & 2 &  $ 2\Phi_{114}-\Phi_{123}-\Phi_{213}  $  \\ \hline
  $H_{g}$   & 1 &  $ \sqrt{3}\ \Phi_{442}               $  \\ 
  $H_{g}$   & 2 &  $ -\Phi_{441}+2\Phi_{234}-\Phi_{324} $  \\
  $H_{g}$   & 3 &  $ \Phi_{332}-2\Phi_{134}+\Phi_{314}  $  \\
  $H_{g}$   & 4 &  $ -\sqrt{3}\ \Phi_{331}              $  \\ \hline
  $H_{u}$   & 1 &  $ \sqrt{3}\ \Phi_{224}                $  \\ 
  $H_{u}$   & 2 &  $ -\Phi_{223}-\Phi_{124}-\Phi_{214}  $  \\
  $H_{u}$   & 3 &  $ \Phi_{114}+\Phi_{123}+\Phi_{213}   $  \\
  $H_{u}$   & 4 &  $ -\sqrt{3}\ \Phi_{113}              $  \\ \hline
\end{tabular}
\end{center}
\end{table}

Although we focused on tri-quark baryon operators, our approach
to constructing hadronic operators is applicable to mesons and
penta-quark operators.  By including $\bm{E}$ and $\bm{B}$ fields
into our operators, hadrons bound by an excited gluon field
can also be studied.
This work was supported by the U.S.~National Science Foundation 
under Awards PHY-0099450 and PHY-0300065, and by 
the U.S.~Department of Energy under
contracts DE-AC05-84ER40150 and DE-FG02-93ER-40762.

\end{document}